\documentstyle[preprint,revtex]{aps}
\begin{document}
\preprint{IFUSP/P-1017}
\draft

\begin{title}
Non-Perturbative Solutions of String Theory \\
in Gravitational Backgrounds
\end{title}

\author{Oct\'{a}vio A. Mattos \cite{email}}
\begin{instit}
 King's College London, Department of Mathematics
\\  The Strand, London WC2R 2LS - U.K.
\end{instit}
\author{Victor O. Rivelles \cite{email1}}
\begin{instit}
Instituto de F\'{\i}sica, Universidade de S\~{a}o Paulo \\
P.O. Box 20516, 01498-970 -- S\~{a}o Paulo - SP - Brazil
\end{instit}

\begin{abstract}
We study bosonic string theory in a gravitational
background. We show that  either left-movers or right-movers
are the only background independent non-perturbative
solutions of the field equations for an arbitrary static
metric. They are stable and have a conserved topological
charge being therefore topological solitons. The action
vanishes for these solutions and hence they provide the
dominating contribution in a path integral quantization.
\end{abstract}


%\narrowtext

The study of non-perturbative solutions in classical field
theories has shed light in many features of the
corresponding quantum theory since in general these features
cannot be reached by perturbation theory. For instance,
topological solitons indicate the existence of hidden
sectors of the Hilbert space while the presence of
instantons indicates vacuum tunnelling, showing, in both
cases, that a rich quantum structure is available. Since
there is no manageable field theory for strings the only way
to study the interaction of strings is through perturbation
theory. Therefore any non-perturbative solution of string
theory may give some clues about the symmetries or the
geometrical foundations of a string field theory.

Another aim of studying strings in backgrounds consists in
finding a (low energy) phenomenologicaly acceptable string
vacuum or consistent compactifications. However, as is well
known, strings do not propagate in arbitrary backgrounds if
conformal symmetry is to be kept at the quantum level
\cite{Callan}.

Since we find background independent solutions
conformal invariance can be achieved by choosing an
appropriate background. Plane waves backgrounds
\cite{Horowitz}, of which shock waves backgrounds
\cite{Amati} are a particular case, as well as group
manifolds \cite{Witten} are examples of backgrounds
which keep conformal invariance at the quantum level.
Recently, a class of space-time metric has been proved
finite in the presence of the dilaton field \cite{Tseytlin}.

In this paper we study non-perturbative solutions of the
bosonic string theory in a gravitational background field.
We show that closed string left-movers or right-movers are
background independent solutions. However, not all
coordinates can be taken as right- or left-movers since in
this case the gauge freedom cannot be fixed completely. In
the light-cone gauge (LCG) we find that only left- or only
right-movers are solutions for static, but otherwise
arbitrary backgrounds, and that these backgrounds can be
interpreted as a comoving frame for the string. When both
movers are present the space-time metric split up in blocks
with two of them being still completely arbitrary.

 In fact, all coordinates being either left- or right-movers
corresponds to the {\em only} background independent
solution. This solution is stable under small perturbations.
Finally we show that the solutions having only the string
winding mode can be interpreted as solitons and a quantum
soliton operator can be built. Since the action vanishes for
left- or right-movers they give the dominating contribution
to the path integral.

The string action in an arbitrary $D$ dimensional
gravitational background $G_{\mu \nu}$ $ (\mu, \nu = 0,
\dots , D-1)$ is given by

\begin{equation}
S=\frac{1}{2\pi \alpha'} \int d\sigma d\tau \sqrt{g} \ g^{
\alpha \beta} G_{\mu \nu}(X)\partial_{\alpha} X^{\mu }
\partial_{\beta}X^{\nu}    ,            \label {eq:acne}
\end{equation}

\noindent
where $g_{\alpha \beta}$ $(\alpha, \beta = 0, 1)$ is the
world-sheet metric. The classical equations of motion and
constraints are, respectively,

\begin{equation}
\Box^{2} X^{\mu } + \Gamma^{\mu }_{\nu \lambda} g^{\alpha
\beta } \partial_{\alpha} X^{\nu} \partial_{ \beta} X^{
\lambda }  =0 ,  \label {eq:mot}       \end{equation}

\FL
\begin{equation}
T_{\alpha \beta } =G_{\mu \nu} (\partial_{\alpha} X^{\mu }
\partial_{ \beta}X^{\nu }-{\scriptstyle \frac{1}{2}}
g_{\alpha \beta } g^{\alpha' \beta' }\partial_{\alpha' }
X^{\mu } \partial_{ \beta' }X^{\nu } )=0   , \label {eq:vin}
\end{equation}

\noindent
where $\Gamma_{\mu \nu}^\lambda$ is the Christoffel symbol
for the metric $G_{\mu \nu}$. In the conformal gauge,
(\ref{eq:mot}) and (\ref{eq:vin}) reduce to

\begin{equation}
\hat{\partial}_+ \hat{\partial}_- X^\mu + \Gamma^\mu_{\nu
\lambda} \hat{\partial}_+ X^\nu \hat{\partial}_- X^\lambda =
0    \label{eq:mot1}        \end{equation}

\begin{equation}
T_{\pm \pm} = G_{\mu \nu} \hat{\partial}_\pm X^\mu
\hat{\partial}_\pm X^\nu = 0
 \label{eq:co}                     \end{equation}
where $\hat{\partial}_\pm \equiv (\partial_\tau \pm
\partial_\sigma )$.

As it is easily seen, left- or right-movers, satisfying $
\hat{\partial}_+ X^\mu = 0$ or $\hat{\partial}_- X^\mu = 0$
respectively, are background independent solutions of the
field equations (\ref{eq:mot1}). The Euclidean version of
these solutions have been recognized as instantons in two-
dimensional non-linear sigma models for target spaces with
non-trivial second homotopy group \cite{Perelomov}. In our
case, however, there are two more equations to be satisfied
namely, the constraints (\ref{eq:co}). As will be shown,
they prove to be very strong, restricting the allowed
background space-times.

In two-dimensional Minkowsky space-time, fields satisfying
either of the conditions $\hat{\partial}_+ X^\mu = 0$ or
$\hat{\partial}_- X^\mu = 0$, are known as chiral bosons.
They have been studied and their quantization performed in
several schemes \cite{Jackiw}. In the heterotic string
theory, left-mover target space coordinates generate the
internal symmetry group through compactification on a flat
torus \cite{Gross}, whereas in the context of non-linear
sigma models, chiral bosonic fields --- the so called
leftons and rightons --- have also been considered
\cite{WS}. We shall denote string coordinates of only a
mover type (either left or right) as being `chiral'.

As it is well known the gauge $g_{\alpha \beta} \! =\!
\eta_{\alpha \beta}$ still leaves a combination of
reparametrization invariance and Weyl scaling to be fixed.
This freedom enables us to make a further gauge choice and
the light-cone gauge $X^{+}\equiv \frac{1}{ \sqrt{2}}
(X^{0}+X^{D-1})=(X_{0}^{+}+P^{+}\tau )$ is taken in order to
fix this residual invariance \cite{GSW}. In this gauge,
however, the coordinate $X^+$ is no longer chiral. We could
therefore ask whether it is possible to obtain a solution
with all coordinates being chiral with a gauge choice such
as $X^{+}=X_{0}^{+}+P^{+} \sigma^{+}$ (where $\sigma^{ \pm}
=\sigma \pm \tau $). Nevertheless, this gauge does not
completely fix the residual gauge freedom left over by the
conformal gauge namely, $\sigma^{+}\rightarrow \tilde{\sigma
}^{+}(\sigma^{ +})$, $\sigma^{-} \rightarrow \tilde{\sigma
}^{-}(\sigma^{-})$, as can be easily checked.

We now turn to study the backgrounds where not all
coordinates are chiral. In order to fix the residual gauge
invariance we choose the LCG with the non-chiral  coordinate
$X^{+}$. The LCG choice will affect the equations of motion
(\ref{eq:mot1}) where all terms containing $X^{+}$ will
spoil the possibility of a chiral solution for the remaining
coordinates. Starting with a metric such that $G_{++}=G_{--}
=0 $, which partially sets the coordinate system
\cite{foot}, we can require that these various terms
containing $X^{+}$ vanish. For right-movers this leads to
the equations

\begin{mathletters}
\FL
\begin{eqnarray}
2P^{+}\partial_{+}G_{+-}+(\partial_{+}G_{-i}+ \partial_{
i}G_{+-}-\partial_{-}G_{+i})\hat{\partial }_{-}X^{i} &=&0 ,
\label {eq:qa1} \\
2P^{+}\partial_{+}G_{+i}+(\partial_{+}G_{-i}+ \partial_{ -
}G_{+i}-\partial_{i}G_{+-})\hat{\partial}_{-}X^{-} &&
\nonumber \\ +(\partial_{+}G_{ij}+ \partial_{[j}G_{+i]})
\hat{\partial}_{-}X^{j}   & =&0 ,   \label {eq:qa2}
\end{eqnarray}

\noindent
where $\partial_{\pm}$ refers to space-time derivatives
(while $\hat{\partial}_{\pm}$ refers to world-sheet
derivatives). The constraints (\ref{eq:co}) then imply,

\begin{eqnarray}
&&2P^{+}G_{+-}\hat{\partial}_{-}X^{-}+2P^{+}G_{+i}\hat{
\partial}_{-}X^{i} \nonumber \\ &+&2G_{-i}\hat{\partial}_{-}
X^{-}\hat{\partial}_{-}X^{i} +G_{ij}\hat{ \partial}_{-}
X^{i}\hat{\partial}_{-}X^{j}  =  0.    \label {eq:qa}
\end{eqnarray}
\end{mathletters}
\label {eq:6}

Now we must solve (\ref{eq:6}) for the various components of
the metric as functions of $X^{+}$, $X^{-}$ and $X^{i}$. It
turns out that these equations have a solution for a static
(i.e, $X^{+}$ independent) background,

\begin{eqnarray}
G_{ij}=g_{ij} \; ,\;\;\;\;\; G_{+i}=\partial_{i}h \; ,
\nonumber \\ G_{+-}=\partial_{-}h \; , \;\;\;\;\; G_{-i}
=g_{-i},  \label {eq:st}          \end{eqnarray}

\noindent
where $g_{ij}$, $g_{-i}$ and $h$ are functions of $X^{-}$
and $X^{i}$ only. Notice that, apart from being static,
(\ref{eq:st}) corresponds to a choice of a coordinate frame,
as the following counting of its independent components
shows. The spatial metric $g_{ij}$ has $\frac{1}{2}(D-2) (D-
1)$ independent components while $g_{-i}$ has $D-2$ and $h$,
one. Altogether they give the $\frac{1}{2}D(D-1)$
independent components of a D-dimensional metric after a
choice of a coordinate frame is made.

The static character of the solution (\ref{eq:st}) stems
from the fact that the chiral coordinates $X^-$ and $X^i$
depend symmetrically on the world-sheet parameters $\sigma $
and $\tau $ while, in the LCG, this symmetry is absent in
$X^+$. Therefore, an equation of motion for the string
coordinates cannot in principle be satisfied once the
background depends on $X^+$, but must be static. It can also
be understood with the following argument. If we start with
a flat space-time where there is a closed string with both
left- and right-movers, then we could go over to a curved
background where only one of the movers remain, according to
our solution. As this cannot happen, it means that it is not
possible to curve the flat space. Therefore we must have a
static background, or at least a background with a time
dependence that can never reach flat space. If we insist in
curving the background we would find a solution with both
movers mixed up non-linearly in order to satisfy the
equations of motion (\ref{eq:mot1}) and the constraints
(\ref{eq:co}).

An important feature of our solution is that the
 center-of-mass momentum is always equal (up to a sign) to
its winding number, as can be seen, for instance, from the
right-mover

\FL
\begin{equation}
X_{R}^{i}(\sigma ,\tau )=x_{R}^{i} +\left( \frac{P^{i}}{2}
-L^{i}\right)(\tau -\sigma )  +\frac{i}{2}\sum_{n\neq 0}
\frac{1}{n}\tilde{\alpha}_{n}e^{-2in(\tau -\sigma )}
\end{equation}

\noindent
where $x_{R}^{i}$ is the zero-mode and, $P^{i}$ and $L^{i}$
are the momenta and winding modes of the associated
non-chiral string which now appear combined to give what is
both the momenta and the winding number of the chiral
string. This makes the string to slip over itself in such a
way that its center-of-mass is always at rest. In fact, the
class of metrics (\ref{eq:st}) corresponds to a comoving
frame for the string.

For a relativistic point particle the comoving frame is
defined by \cite{Weinberg}

\begin{equation}
ds^{2} =dt^{2} -G_{ij}(X,t)dX^{i}dX^{j}.
\label {eq:com}         \end{equation}
Once we go to the proper-time gauge ${X}^{0} =P^{0} \tau $
(it plays a role similar to the conformal and the LCG for
the string), the equations of motion for the point particle
take the form

\begin{mathletters}
\begin{equation}
\ddot{X}^{i}+G^{ik}\dot{G}_{kj}\dot{X}^{j} -\frac{1}{2}
G^{ij}\partial_{j}G_{kl}\dot{X}^{k}\dot{X}^{l} =0      ;
\label {eq:ll}               \end{equation}

\noindent
and the point particle constraint, $P^{2}=m^{2}$, becomes,

\begin{equation}
(P^{0})^{2}-G_{ij}\dot{X}^{i}\dot{X}^{j}= m^{2},
\label {eq:ll1}         \end{equation}
\end{mathletters}
where $m$ is the particle rest mass. The only background
independent solution to these equations is $\dot{X}^{i}=0$.
To see this take the partial derivative of the constraint
(\ref{eq:ll1}),

\begin{equation}
(\partial_{k}G_{ij})\dot{X}^{i}\dot{X}^{j}=0 .
\end{equation}
Thus, (\ref{eq:ll}) gives, $\partial_{\tau}(G_{ij} \dot{X
}^{j}) =0$, or $\dot{X}^{t}=G^{jk}C_{k}$, for a constant $
C_{k}$. Substituting in (\ref{eq:ll1}), gives $C_{k} C_{l}
G^{kl}=cte $. Since $G_{kl}$ is arbitrary, this identity can
only be satisfied if $C_{k}=0$, or $\dot{X }^{i}=0$. The
equation of motion for $X^{0}$ (namely $\dot{X}^{i}\dot{X}^{
j} \partial_{0}G_{ij} =0$) is consistent with (\ref{eq:ll1})
and was not directly used in this derivation.

For the string, it can be easily demonstrated, with a metric
of the form

\begin{equation}
\begin{array}{cc}
G_{\mu \nu}(X^{i}) = & \left( \begin{array}{cc}
\left( \begin{array}{cc} 0 & 1 \\ 1 & 0 \end{array} \right)
& 0  \\ 0  & G_{ij} \end{array} \right),
\end{array} \label {eq:me} \end{equation}
that either left- or right-movers are the unique background
independent solutions.

\noindent
The equation of motion for $X^{+}$ gives $\partial_{-}
G_{ij}=0 $, while the equation of motion for $X^{-}$ (which
is the same as for the flat case), for the $X^{i}$, and the
constraints are, respectively,

\begin{mathletters}
\begin{equation}
\hat{\partial}_{+}\hat{\partial}_{-}X^{-}=0  ,
\label {eq:ll3} \end{equation}
\FL
\begin{equation}
\hat{\partial}_{+}\hat{\partial}_{-} X^{i} +\frac{1}{2}
G^{il}(\partial_{j}G_{kl}+\partial_{k}G_{jl}-\partial_{l}
G_{jk}) \hat{\partial}_{+}X^{i}\hat{\partial}_{-}X^{k}=0 ,
\label {eq:ll5}         \end{equation}

\begin{equation}
2P_{+}\hat{\partial}_{\pm}X^{-}=G_{ij}\hat{\partial}_{\pm }
X^{i} \hat{\partial}_{\pm }X^{j}  .
\label {eq:ll4} \end{equation}
\end{mathletters}
{}From (\ref{eq:ll3}) and (\ref{eq:ll4}) one has

\begin{equation}
G_{ij}\hat{\partial}_{\pm}X^{i}\hat{\partial}_{\pm}X^{j}=F(\
sigma^{\pm}).            \label {eq:14}          \end{equation}
This last equation allow us in fact to have both movers. To
take this into account let us split the coordinates $X^i$
into left-movers coordinates $X^A, (\hat{\partial}_{-} X^{
A}=0, A=1,...,D'-1)$ and right movers $X^a, (\hat{\partial
}_{+} X^{a}=0, a=D',...,D-2)$. Without loss of generality,
the components of the metric $G_{ij}$ can also be split into
$ G_{ab}, \; G_{AB}$ and $G_{aB}$. Then from (\ref{eq:14})
we obtain $G_{AB}$ as an arbitrary function of $X^A$ and
similarly for $G_{ab}$, while from (\ref{eq:ll5}) one gets,

\begin{eqnarray}
(\partial_{B}G_{Ab}-\partial_{A}G_{Bb})\hat{\partial}_{-
}X^{b} \hat{\partial}_{+}X^{B}&=&0 \;\;\;\;\;\;\;\;\;\;
\mbox{for i=A}, \nonumber \\
(\partial_{b}G_{aB}-\partial_{ a}G_{bB})\hat{\partial}_{-
}X^{b} \hat{\partial}_{+}X^{B}&=&0 \;\;\;\;\;\;\;\;\;\;
\mbox{for i=a}
\end{eqnarray}
A local solution for the mixed part of the metric is
$G_{aB}=\partial_{a} \partial_{B}f$ where $f$ is an
arbitrary function of $X^A$ and $X^a$. The background
allowing both movers is then

\begin{eqnarray}
G_{AB} = G_{AB}(X^A) \; , \;\;\;\; G_{ab} = G_{ab}(X^a) \; ,
\nonumber \\ G_{aB}=\partial_{a}\partial_{B} f(X^A, X^a).
\label {eq:aB} \end{eqnarray}

By substituting (\ref{eq:aB}) in the equations of motion one
does indeed see that this is the most general class of
metric which allows the presence of both movers.
Nevertheless, for an {\em arbitrary} metric $G_{ij}$, {\em
only} the solution with one of the movers present is
allowed. For metrics of the form (\ref{eq:st}) the chiral
solution is {\em also} unique. This stems from the fact that
an arbitrary background of type (\ref{eq:st}) is more
general than an arbitrary background of type (\ref{eq:me}).

Now we turn to the study of the stability of the solution.
Since our solution can be regarded as, say, a right-mover
closed string with its center-of-mass at rest in a comoving
frame we have to look for small perturbations with can grow
up with time $\tau$ or small perturbations which can change
the chirality of the string. The question of stability can
then be posed as a problem of finding the eigenvalues $w$
for a perturbation $f(\sigma)$ with small coefficient
$\lambda $,

\begin{equation}
X^{i}(\sigma ,\tau )=X^{i}(\sigma -\tau ) +\lambda
f^{i}_{w}(\sigma )e^{-iw \tau } .       \label {eq: }
\end{equation}
As usual, real eigenvalues indicates the stability of the
solution. By substituting it into the equations of motion
and keeping only the linear terms it can then be easily seen
that, due to the dependence on $\tau $ of the Christoffel
symbol (and coordinate $X^{j}$),
\FL
\begin{equation}
f''^{i}(\sigma ) +w^{2}f^{i}(\sigma )=-2[\Gamma^{i}_{jk}
\dot{X}^{j}](\sigma ,\tau ) [iwf^{k}(\sigma )-f'^{k}(\sigma
)],      \label {eq:ei} \end{equation}

\noindent
the eigenvalue equations decouple and we get $f(\sigma)=e^{
iw \sigma }$ with $w$ real. Then the perturbation does not
change the character of the solution (it remains a right-
mover) and we  conclude that the solution is stable under
small perturbations.

Besides having zero action, which means a large contribution
to the path integral, another important feature of the
chiral solution in curved space is its solitonic character.
The modes of the string are not affected by the presence of
the background, a behaviour that resembles solitons. The
same occurs with the winding mode, which, in addition, has a
topological character. On the other hand, the Hamiltonian

\begin{equation}
H= \int d\sigma \ G_{ij}\dot{X}^{i}\dot{X}^{j} =\int d\sigma
\ G_{ij} X'^{i}X'^{j}    ,    \end{equation}

\noindent
does depend on the metric. This can be interpreted even if
the string is understood as being in a comoving frame, since
it is an extended object and its energy is not localized.

As in flat space-time \cite{Ezawa} the string winding mode
has a solitonic character. Since we have a topological
conserved current $J^{i}_{\alpha }=(2\pi) \epsilon_{\alpha
\beta }\partial^{\beta }X^{i}$, there is a conserved charge
for a solitonic solution of the form $X_{sol}^{i}( \sigma -
\tau )=2w^{i}(\sigma -\tau )$, where $w^i$ is the winding
number. A formal expression for the quantum soliton operator
which creates a soliton (of winding number $w^i$) out of the
vacuum is given by

\begin{equation}
T(w) =\exp [-\frac{i}{\pi }\int d \sigma^{-} \ X_{sol}^{i}(
\sigma -\tau ) P_{i}(\sigma -\tau )];
\end{equation}

\noindent
where $P_i$ is the momentum associated to $X^i$, i.e,
$P_{i}=G_{ij} \hat{\partial}_{-}X^{j}$. This operator
satisfies the usual commutation rule with the charge
operator $\hat{W}^{i}=(2\pi )^{-1}\int d \sigma^{-} \
\hat{\partial}_{-} X^{i}$, that is,

\begin{equation}
[\hat{W}^{i}, T(w)]=w^{i}T(w).
\end{equation}

We conclude with some comments and remarks. As far as the
properties of the string solution  are concerned, only local
properties of the space-time, like its curvature, are
relevant. Topological properties such as a possible
nonsimply connected structure or compactedness of the
background, are not relevant to select one of the movers.
The later properties may contribute to a nontrivial
fundamental homotopy group which in turn determines the
presence of winding modes only. In connection with this,
other backgrounds such as the antisymmetric tensor can be
considered. It will contribute with a torsion term to the
equations of motion.

The solution allowing both movers eq. (\ref{eq:aB}) could be
turned into a K\"ahler metric if $G_{AB} = G_{ab} = 0$ and
if a complex structure is properly introduced after
Euclideanization. The left- or right-movers turn into
holomorphic or antiholomorphic functions and one may
interpret them as sigma-model instantons defined on a
K\"ahler manifold \cite{Perelomov1}. However this
interpretation is not valid since our base space is compact
(in the world sheet coordinate $\sigma$) and it is not
possible, by joining a point at infinity to define the
second homotopy group needed for the characterisation of
instantons.

Besides the static backgrounds which allows left- and/or
right-movers, we also found from eqs. (\ref{eq:qa1}),
(\ref{eq:qa2}) and (\ref{eq:qa}) a time-dependent solution

\begin{mathletters}
\begin{eqnarray}
G_{+-}&=&\partial_{-}h \\
G_{-i} &=& [a_{i}X^{+}+g_{i}(X^{i})]G_{+-} \\
G_{+i}&=&a_{i}h+\partial_{i}h \\
G_{ij}&=&\frac{G_{+(i}G_{-j)}}{G_{+-}}     ,
  \end{eqnarray}        \end{mathletters}
\label {eq:sing}

\noindent
where the $a_{i}$ are constants, $g_{i}=g_{i}(X^{i})$ and
$h=h(X^{-},X^{i})$. As pointed out above such a
time-dependent solution should not be allowed. In fact, it
is possible to show that the metric (\ref{eq:sing}) is
singular. Singular metrics are allowed in gauge theories for
gravitation but their role is not understood \cite{bh}.
 Moreover it should be noticed that such a metric admits
black hole solutions since, if $G_{+-}$ vanishes in some
point, then $G_{ij}$ becomes singular (and
vice-versa).

Other background fields may be considered. The antisymmetric
tensor, for instance, contributes to the equations of motion
with a torsion which is the antisymmetric part of the
Christoffel symbol. This generalized equation still allows
the chiral solution.

\acknowledgments

OAM would like to thank the support received at the
Instituto de F\'{\i}sica of the Universidade de S\~{a}o
Paulo, and CAPES for financial support. The work of VOR is
partially supported by CNPq.


\begin{references}
\bibitem{foot} A space $M^{2}\oplus K$, with $M^{2}$ being
the 2-dimensional Minkowski space and K some curved space
(not necessarily compact), enables through the LCG to
determine the $X^{-}$ coordinate algebraically in terms of
transverse coordinates. With our choice for the metric one
can still determine $X^{-}$ by solving a differential
equation.
\end{references}
\end{document}